\documentclass[prl,twocolumn]{revtex4}  		
\usepackage{graphicx}

\begin{document}

\title
{frustration-induced phase transitions in 
the spin-$S$ orthogonal-dimer chain}

\author
{   							
Akihisa Koga and Norio Kawakami
}

\affiliation
{
Department of Applied Physics, Osaka University, Suita, Osaka 565-0871, Japan
}

\date{\today}

\begin{abstract}
We investigate quantum phase transitions in the 
frustrated orthogonal-dimer chain with an arbitrary spin $S \geq 1/2$.
When the ratio of the competing exchange couplings is varied, 
first-order phase transitions occur $2S$ times among distinct
spin-gap phases.  The introduction of 
 single-ion anisotropy further enriches the phase diagram.
The phase transitions described by the present model possess most of
the essential properties inherent in frustrated  quantum spin systems.
\end{abstract}

\pacs{Valid PACS appear here}%

\maketitle

Geometrical frustration in strongly correlated electron systems 
has attracted much current interest. 
A remarkable example of materials is the two-dimensional spin-gap
compound $\rm SrCu_2(BO_3)_2$,\cite{Kageyama}
in which the characteristic orthogonal-dimer
 structure of the $\rm Cu^{2+}$ ions
stabilizes the spin-singlet ground state.\cite{Shastry,Miyahara,SrTheory}
Remarkably,  strong frustration induces
 an anomalous first-order phase transition
 in addition to the plateau-formation 
in the magnetization  process.\cite{Onizuka} 
More recently, another orthogonal-dimer
compound $\rm Nd_2BaZnO_5$\cite{Kageyama2} 
was synthesized, where higher-spin and orbital moments ($J=9/2$)
show the antiferromagnetic order
at the critical temperature $T_{\rm N}=2.4{\rm K}$,
making such higher-spin systems more interesting.

One of the most prototypical phenomena inherent in 
the frustrated systems is the {\it first}-order transition,
which is triggered by the competition of various phases
due to strong frustration. Although the systematic treatment of
first-order transitions is not easy, it is highly desirable to
clarify  how such first-order transitions are 
induced by frustration in order to understand 
the essential properties common to such
frustrated quantum spin systems.

In this paper, we investigate a remarkable one-dimensional (1D)
spin-$S$ orthogonal-dimer model, which possesses most of the
essential properties of first-order
phase transitions in this class of frustrated 
spin systems. \cite{Kato,Ivanov,plachn,Takushima}
By exploiting the non-linear sigma model (NL$\sigma$M) approach as well as
the exact diagonalization and the series expansion,
we find the $(2S+1)$ distinct spin-gap phases in 
the spin-$S$ chain, 
which are separated by first-order quantum phase transitions.
The effect of single-ion anisotropy is also addressed.
We demonstrate that a higher-spin generalization
of the model results in the remarkably rich phase diagram, which realizes
the idea of valence-bond-solid (VBS)  \cite{VBS} 
in {\it a sequence of first-order phase transitions}.

\begin{figure}[htb]
\begin{center}
\includegraphics[width=7cm]{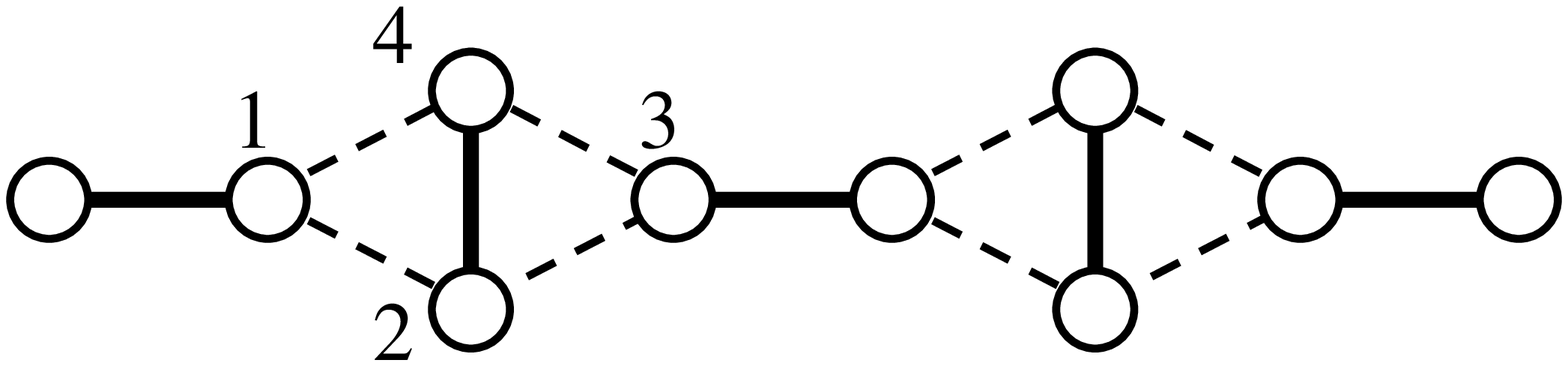}
\end{center}
\vskip -4mm
\caption{Orthogonal-dimer spin chain: the dimer
bonds shown by solid lines have the characteristic orthogonal
structure.
}
\label{fig:model1}
\end{figure}

Let us consider the 1D quantum spin system with
the orthogonal-dimer structure \cite{Kato,Ivanov,plachn}
 shown schematically in Fig. \ref{fig:model1}.
The corresponding Hamiltonian reads
\begin{eqnarray}
H&=&J\sum_{j}\left({\bf S}_{2,j}\cdot{\bf S}_{4,j}+
{\bf S}_{3,j}\cdot{\bf S}_{1,j+1}
\right)\nonumber
\\&&+J'\sum_j\left({\bf S}_{2,j}+{\bf S}_{4,j}\right)\cdot
\left({\bf S}_{1,j}+{\bf S}_{3,j}\right),\label{eq:H}
\end{eqnarray}
where ${\bf S}_{i,j}$ is the $i$-th spin operator in the $j$-th plaquette
and $J (J')$ is the antiferromagnetic exchange coupling.
The $S=1/2$ model  was so far studied well.\cite{Ivanov,plachn}
 Below, we focus on a higher-spin generalization of 
the system.

We first exploit the NL$\sigma$M technique
to clarify the topological nature
of the system.\cite{Haldane,Affleck,Fukui,first,Takano}
Note that the Hamiltonian has 
the remarkable relation 
$\left[H,{\bf S}_{2,j}+{\bf S}_{4,j}\right]=0$.
Therefore, by introducing the composite spin $T_j$ defined as 
${\bf T}_j={\bf S}_{2,j}+{\bf S}_{4,j}$,
we obtain the effective mixed-spin Hamiltonian as,
$H=\sum\left[J'{\bf S}_{1j}\cdot{\bf T}_j+J'{\bf T}_j\cdot{\bf S}_{3j}+
J{\bf S}_{3j}\cdot{\bf S}_{1j+1}\right]
+\frac{J}{2}\sum T_j(T_j+1)-\frac{1}{4}JS(S+1)N$,
where $N$ is the total number of sites.
Then, the Hilbert space of the Hamiltonian (\ref{eq:H}) 
can be classified into each sub-space specified by $[S; \{ T_j\}]$.
The singlet ground state as well as relevant low-energy excitations
are in the space with uniform $T_j$ ($=T$). In particular,
for a given  $T\neq 0$ we can describe 
low-energy properties by the NL$\sigma$M ($T=0$
gives a trivial system with decoupled dimers).
Introducing three kinds of the fluctuation 
fields,\cite{Fukui,first,Takano}
we obtain the Euclidean Lagrangian ${\cal L}$
with the effective field ${\bf \phi}$ as, 
\begin{eqnarray}
{\cal L}&=&\frac{1}{2g}
\left(v_s{\bf \phi}^{'2}+\frac{1}{v_s}\dot{{\bf \phi}}^2\right) 	   
-\frac{i\theta}{4\pi}{\bf \phi}\cdot({\bf \phi}^{\prime}\times\dot{\bf \phi}),
\end{eqnarray}  
where $\theta=2\pi T$, $g=2AB/T$, $v_s=6J'StB/A$ with $A=(2t+j_0)^{1/2}$
$B=[4j_0t^2+2(1-2j_0)t+j_0]^{-1/2}$,  $j_0=J'/J$ and $t=S/T$.
Note that the topological angle $\theta$ is zero (mod $2\pi$) 
for any choice of $T$, and the system is always gapped. 
Therefore, in case the phase transition between 
the distinct sub-spaces $[S; \{T_j\}]$ occurs,
it should be accompanied by the discontinuity 
in the parameters $g$ and $v_s$.
This suggests that the possible quantum phase transition 
should  be of {\it first}-order.

\begin{figure}[htb]
\begin{center}
\includegraphics[width=7cm]{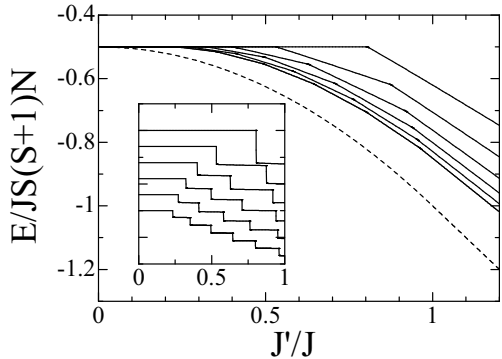}
\end{center}
\vskip -4mm
\caption{Ground-state energy as a function of $J'/J$
obtained by the exact diagonalization for 8 sites.
From up to down $S=1/2, 1, 3/2, 2, 5/2, 3$.
The broken line is the energy for the classical limit 
$S\rightarrow\infty$.
The inset shows the derivative of the ground state energy, 
in which the origin for each curve is shifted for convenience.
}
\label{fig:eg}
\end{figure}

\begin{figure}[htb]
\begin{center}
\includegraphics[width=7cm]{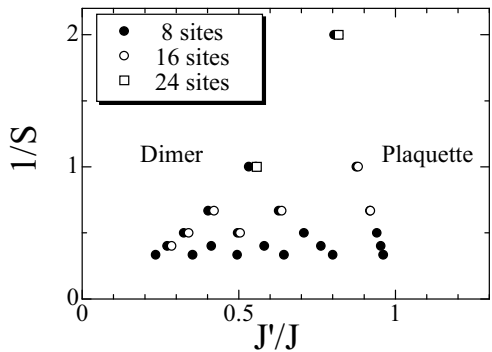}
\vskip -4mm
\end{center}
\caption{Ground-state phase diagram for the orthogonal-dimer spin chain
with generic spin $S$.
The closed circles, open circles and open squares indicate 
the phase boundaries determined by the spin chain $(N=8,16,24)$ 
with periodic boundary conditions.
}
\label{fig:phase}
\end{figure}

To clarify this point, we numerically diagonalize the Hamiltonian
(\ref{eq:H}) for a small cluster with periodic boundary conditions.
The ground state energy is shown in Fig. \ref{fig:eg}.
It is found that the cusps  appear $2S$ times 
in the energy diagram for the spin-$S$ case,
implying that the {\it first}-order quantum phase transitions 
indeed occur $2S$ times. Also shown is
 the energy in the classical limit $S\rightarrow\infty$,
which is given as $E/JS^2N=-\frac{1}{2}[1+(J'/J)^2]$, and $-J'/J$ 
for the helical phase $(0<J'/J<1)$ 
and the antiferromagnetically ordered phase $(1<J'/J)$.
As $S$ increases, the profile of the 
energy diagram gradually approaches the classical one, 
although  the $2S$ transitions should exist for any finite $S$.
By examining several systems with different cluster size, 
we determine the phase diagram rather precisely,
as shown in Fig. \ref{fig:phase}.
In this way, the present model is the remarkable
spin-$S$ microscopic model, 
which clearly determines a sequence of the phase transitions
triggered by frustration. Moreover, the {\it first}-order
 transitions found here are contrasted to the
{\it second}-order ones known 
for the ordinary spin-$S$ Heisenberg chain with 
bond alternation.\cite{Affleck} 

\begin{figure}[htb]
\begin{center}
\includegraphics[width=7cm]{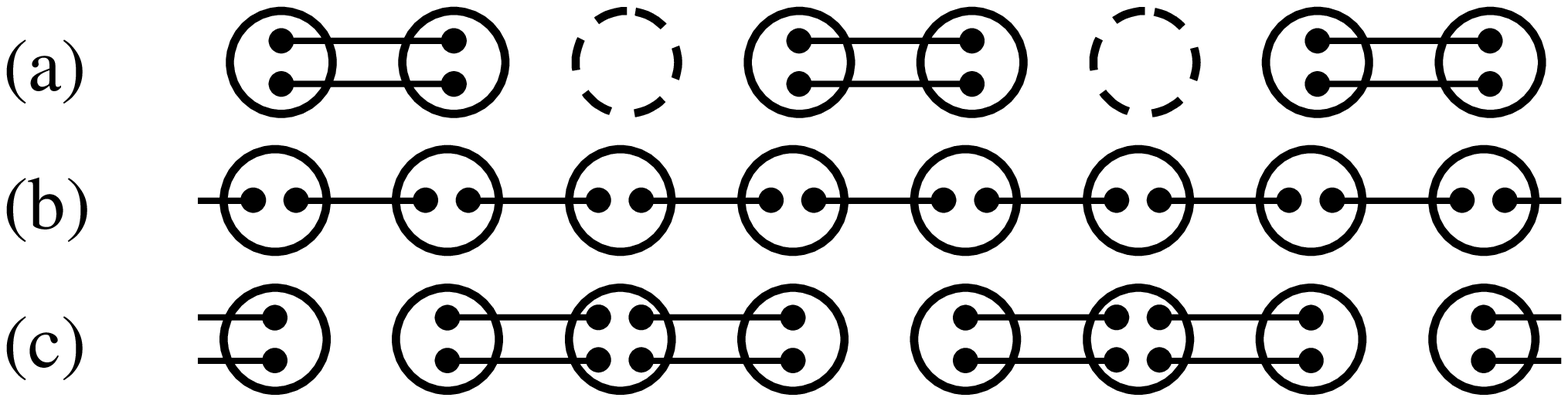}
\end{center}
\vskip -4mm
\caption{VBS picture of the phases for the $S=1$ model.}
\label{fig:VBS}
\end{figure}

We now clarify the nature of 
each spin-gap phase by taking the $S=1$ model 
as an example, which contains three distinct spin-gap phases.
For this purpose, the VBS description \cite{VBS} of
the ground state is useful, 
where the topological nature is specified by
a combination of singlet bonds 
between the decomposed $S=1/2$ spins, as shown 
in Fig. \ref{fig:VBS}.
Recall that in the small (large) $J'/J$ region, the 
composite spin $T=0 (2)$ on each diagonal bond is realized.
Hence the dimer (plaquette) phase characterized by 
Fig. \ref{fig:VBS} (a) [(c)] is stabilized there.
On the other hand, in the intermediate phase, 
strong frustration induces the spin $T=1$ on each diagonal bond,
resulting in  the singlet phase characterized 
by Fig. \ref{fig:VBS}(b). This may be regarded as the 
{\it frustration-induced Haldane phase}.

To confirm the above predictions based on the VBS analysis,
we perform the numerical diagonalization of small clusters
in the corresponding sub-spaces.
\begin{figure}[htb]
\begin{center}
\includegraphics[width=7cm]{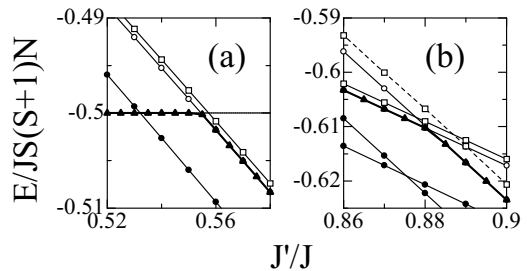}
\end{center}
\vskip -4mm
\caption{The energy for the $S=1$ orthogonal-dimer spin chain
in  the restricted sub-space.
}
\label{fig:s1eg}
\end{figure}
In Fig. \ref{fig:s1eg} (a), 
the flat line 
is the energy for the exact dimer state.
The energy for the Haldane phase ($N=8, 16$ and $24)$
obtained by the exact diagonalization 
in the sub-space $[\{T_j=1\}]$ is shown as the solid line 
with closed circles, open circles and open squares, respectively.
We also show the ground state energy of the original
orthogonal-dimer spin chain 
$(N=16)$ as the bold line with closed triangles.
As seen clearly, the increase of $J'/J$ triggers the first-order quantum 
phase transition from the dimer phase $(T_j=0)$ to the Haldane phase $(T_j=1)$
at the critical value $(J'/J)_c\sim0.56$. Shown in Fig. \ref{fig:s1eg} (b)
is  the energy around the other  first-order  critical point 
between the Haldane phase $(T_j=1)$ and the plaquette phase $(T_j=2)$.
In the plaquette phase, we have used the series expansion 
\cite{series}
by choosing an isolated plaquette as the unperturbed system
and regarding the interaction between plaquettes as a perturbation. 
The ground state energy calculated up to the ninth order 
is shown as the broken line in Fig. \ref{fig:s1eg}, from which
we find the critical point $(J'/J)_c\sim0.88$ between the Haldane and 
the plaquette phases.
It is thus concluded that the first-order quantum phase transitions 
occur among three singlet phases specified by the
distinct sub-spaces $[S; \{T_j\}] (T_j=0, 1, 2)$,
as predicted by the NL$\sigma$M approach.

Keeping this in mind,
we now consider excitations in the $S=1$ orthogonal-dimer spin chain.    
\begin{figure}[htb]
\begin{center}
\includegraphics[width=7cm]{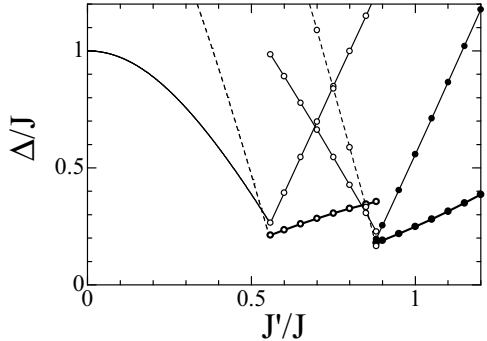}
\end{center}
\vskip -4mm
\caption{Various excitations in the $S=1$ orthogonal-dimer spin chain.
The bold lines indicate a dispersive magnetic excitation, while
the solid (broken) lines  dispersionless magnetic 
(non-magnetic) excitations.
}
\label{fig:excite}
\end{figure}
In the dimer phase $(J'/J<0.56)$, the ground state is given by the product
of isolated dimers $[\{T_j=0\}]$, which allows us to
estimate the excitation gap exactly from the
finite-size calculation. The lowest magnetic excitation 
is described by a defect in the uniform spin alignment, {\it i.e.},
it is given by 
the lowest triplet state in the space of $[1, 0, 0, 0, \cdots]$.
It is also found that in the vicinity of the critical point
($J'/J \sim 0.56$),  a non-magnetic excitation 
belonging to the space $[1, 1, 0, 0, 0,\cdots]$ can be the lowest one,
as shown by the broken line.  In the Haldane phase
($0.56<J'/J <0.88$),
we estimate several kinds of spin 
gaps by extrapolating the results for $N=16, 24$
in the formula  $\Delta(N)=\Delta(\infty)+aN^{-2}$.
The Haldane gap expected naively is the 
lowest away from the critical points,
shown as the bold line in Fig. \ref{fig:excite}.
There are other dispersionless excitations,
which can be described by a defect in the spin alignment such as 
$[0, 1, 1, 1, \cdots]$, $[2, 1, 1, 1, \cdots]$.
These excitations are  bound into 
another non-magnetic state, which is the lowest excitation 
in the Haldane phase close to the plaquette phase
($J'/J \sim 0.88$). In the plaquette phase
($0.88<J'/J$), the series expansion is more efficient to obtain
the dispersive and the dispersionless excitations.
The results computed up to the seventh and 
the ninth order are shown as the closed 
circles in Fig. \ref{fig:excite}.   In this way, several different
excitations become almost degenerate around the 
first-order phase transition points,
reflecting strong frustration.

\begin{figure}[htb]
\begin{center}
\includegraphics[width=7cm]{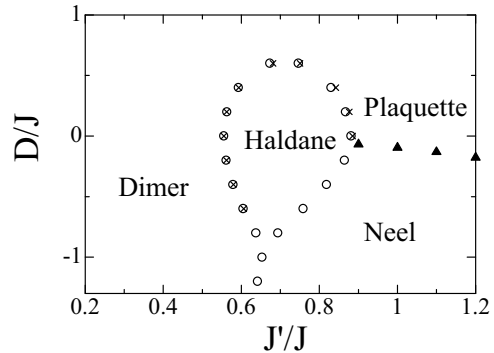}
\end{center}
\vskip -4mm
\caption{Phase diagram for the anisotropic 
 $S=1$ chain with the orthogonal-dimer 
structure. 
}
\label{fig:D-term}
\end{figure}

\begin{figure}[htb]
\begin{center}
\includegraphics[width=7cm]{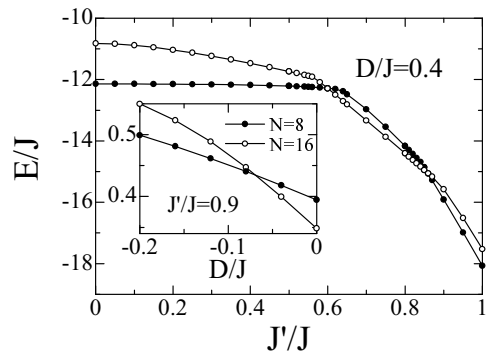}
\end{center}
\vskip -4mm
\caption{Two lowest energies with twisted boundary conditions
for $N=16$ and $D=0.4$. 
The energy of the Haldane state $(P=\tau=-1)$ and the dimer or 
the plaquette state $(P=\tau=1)$ are shown
by the open and closed circles. 		
The inset shows the Binder parameter $U_{Neel}$ for $J'/J=0.9$.
The closed and open circles  ($N=8$ and $16$) 
are calculated with periodic boundary conditions.
}
\label{fig:level}
\end{figure}

Finally we discuss the effect of single-ion anisotropy 
with the Hamiltonian $H_D=D\sum \left(S_{i,j}^z\right)^2$.
The composite spin $T_j$ is not a good quantum number in the
presence of the $D$-term.
The frustration-induced intermediate phases, however, 
are stable against the introduction of small anisotropy.
In fact, the cusps still exist in the energy diagram
for the spin $S=1$ chain $(N=16)$,
from which we determine the phase boundaries 
shown as the open circles in Fig. \ref{fig:D-term}.
The Haldane phase gradually shrinks with the increase of $D(>0)$, 
and finally disappears.
We note that around $D\gtrsim 0.8$ the dimer and the plaquette phases 
indeed merge into the single phase.
In order to clearly 
distinguish the Haldane phase and the dimer (or plaquette) phase,
we make use of the symmetry of the space inversion $P$ and 
the spin reversal $\tau$.
Under twisted boundary conditions\cite{Kitazawa}
$(S_{1,N/4+1}^x=-S_{1,1}^x, S_{1,N/4+1}^y=-S_{1,1}^y, 
S_{1,N/4+1}^z=S_{1,1}^z)$, 
the dimer or plaquette state has 
the eigenvalues $P=\tau=1$, while the Haldane state $P=\tau=-1$.  	   
Therefore, we can distinguish these phases by diagonalizing 
the Hamiltonian with twisted boundary in the restricted space specified by
$P=\tau= \pm 1$.
The results for $D=0.4$ are shown in Fig. \ref{fig:level}.
It is found that the  two lowest energy levels intersect
each other twice when the ratio of 
the exchange coupling $J'/J$ is varied.
This implies that the quantum phase transitions between the phases 
with distinct symmetry $(P=\tau=\pm 1)$ occur twice, 
being consistent with the above predictions based on the 
VBS analysis.
By scaling the critical values as $(J'/J)_c(N)=(J'/J)_c(\infty)+aN^{-2}$,
we estimate the phase boundaries shown as the crosses 
in Fig. \ref{fig:D-term}, which agree well
 with those determined from the original plaquette 
chain $(N=16)$ shown as open circles. In contrast,
when $D$ is negative, the antiferromagnetic correlation 
is enhanced, and the systems is driven to the Neel ordered phase.
To characterize this transition, we check the behavior of
the Binder parameter, $U_{Neel}=1-\frac{<O^4>}{3<O^2>^2}$, 
where $O$ is the order parameter defined as $O=\sum (-1)^{i+j}S_{i,j}^z$.
In the inset of Fig. \ref{fig:level}, 
the Binder parameters for $N=8$ and $16$ are shown
as a function of $D$ with $J'/J=0.9$.
Since the Binder parameter stays invariant with 
the change of $N$ at the transition point,
we can determine the critical value $(D/J)_c\sim -0.07$.
Consequently, we end up with the phase diagram for the spin $S=1$ chain
with anisotropy as shown in Fig. \ref{fig:D-term}.  			

\begin{figure}[htb]
\begin{center}
\includegraphics[width=7cm]{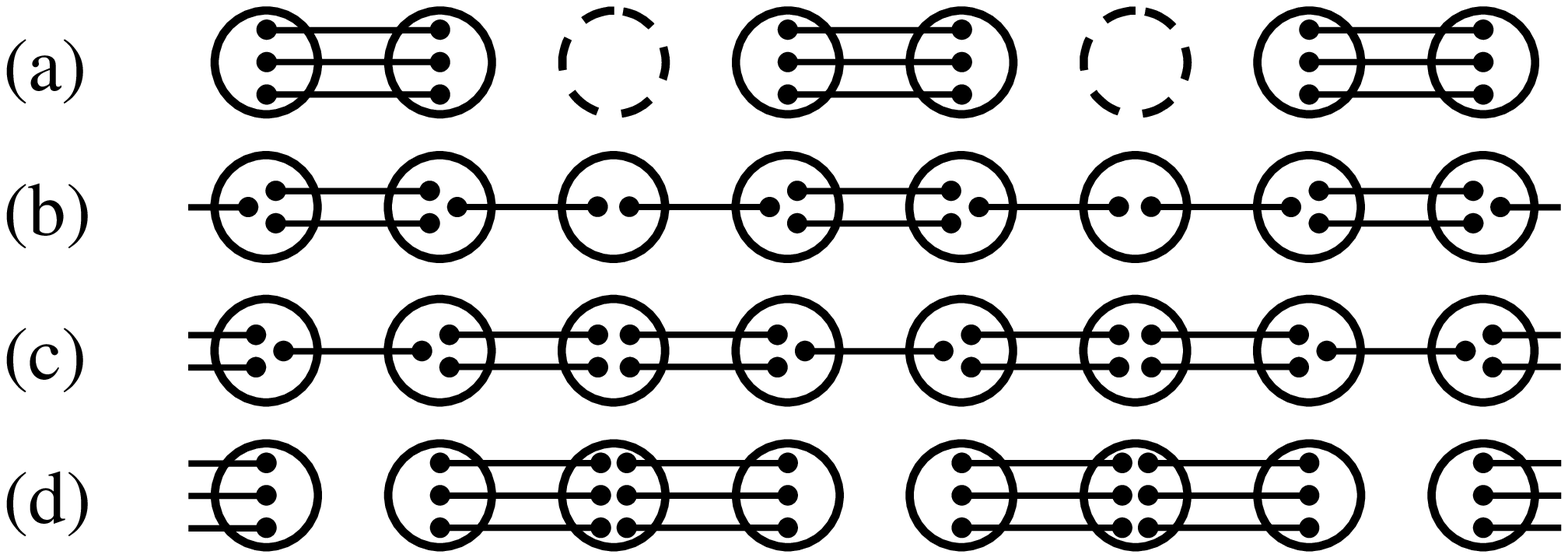}
\end{center}
\vskip -4mm
\caption{VBS picture of the phases for the $S=3/2$ model.}
\label{fig:VBS1.5}
\end{figure}

In summary, we have studied the spin-$S$ quantum spin chain
with the orthogonal-dimer structure.				
The obtained phase diagram has a rich structure
with various type of spin-gap phases which are 
separated by first-order phase transitions 
inherent in fully frustrated systems.
  Although we have given detailed accounts for the
$S=1$ case, it is straightforward to generalize the
discussions to an arbitrary-spin case. For example,
four distinct spin-gap phases for $S=3/2$  in 
Fig. \ref{fig:phase} are completely classified by the VBS states
shown in Fig. \ref{fig:VBS1.5}.  Finally we emphasize that
the first-order phase transitions in the present model
are triggered by strong frustration,
which may capture most of essential properties common to
this class of fully frustrated systems. In particular, we think that
frustration-induced spin-gap phases obtained here
may play a key role to clarify 
the phase diagram of the 2D spin-$S$ orthogonal-dimer model.  

This work was partly supported by a Grant-in-Aid from the Ministry 
of Education, Science, Sports and Culture of Japan. 
A part of computations was done at the Supercomputer Center at the 
Institute for Solid State Physics, University of Tokyo
and Yukawa Institute Computer Facility.

%


\end{document}